\documentclass{ws-ijmpd}

\begin{document}

\markboth{A. G. Alaverdyan, G. B. Alaverdyan and A. O. Chiladze}
{Deconfinement Phase Transition in Neutron Stars and
$\delta$-Meson Field}

\catchline{}{}{}{}{}
%
%%%%%%%%%%%%%%%%%%%%%%%%%%%%%%%%%%%%%%%%%%%%%%%%%%%%%%%%%%%%%%%%%%%%

\title{DECONFINEMENT PHASE TRANSITION IN NEUTRON STARS AND $\delta$-MESON FIELD IN RMF THEORY}

\author{A. G. ALAVERDYAN, G. B. ALAVERDYAN{$\dag$} and A. O. CHILADZE}

\address{Faculty of RadioPhysics, Yerevan State University, Manoogyan str.
1\\ Yerevan, 0025, Armenia\\  {\dag} galaverdyan@ysu.am}

\maketitle

\begin{history}
\received{Day Month Year}
\revised{Day Month Year}
\comby{Managing Editor}
\end{history}

\begin{abstract}
The Maxwell and Glendenning construction scenarios of
deconfinement phase transition in neutron star matter are
investigated. The hadronic phase is described within the
relativistic mean-field (RMF) theory, if also the scalar-isovector
$\delta$-meson field is taken into account. The strange quark
phase is described in the frame of MIT bag model, including the
effect of perturbative one-gluon exchange interactions. The
influence of the $\delta$-meson field on the deconfinement phase
transition boundary characteristics is discussed.
\end{abstract}


\section{Introduction}

The modern concept of hadron-quark phase transition is based on
the feature of such transition that there are two conserved
quantities in this transition: the baryon number and the electric
charge\cite{Gl92}. It is known, that depending on the value of
surface tension, $\sigma_{s}$, the phase transition of nuclear
matter into quark matter can occur in two scenarios\cite{Heis}:
either the ordinary first order phase transition with a density
jump (Maxwell construction-MC), or formation of a mixed
hadron-quark matter with a continuous variation of pressure and
density (Glendenning construction-GC)\cite{Gl92}. The uncertainty
of the surface tension values does not allow to determine the
phase transition scenario, which takes place in realty. In this
paper we investigate the influence of $\delta$-meson effective
field on the phase transition boundary characteristics and compare
two alternative deconfinement phase transition scenarios.

\section{Glendenning and Maxwell constructions for deconfinement phase transition.}
 In order to study the deconfinement phase transition it is
necessary to have the models describing the hadronic matter and
the quark matter. The relativistic mean-field (RMF)
theory\cite{SW} has been effectively applied to describe the EOS
of neutron and protoneutron star matter\cite{MarMil}, as well as
the structure of finite nuclei. For hadronic phase we use the
relativistic Lagrangian density of the many-particle system
consisted in nucleons, $p$, $n$, and exchanged mesons
$\sigma,~\omega,~\rho,~\delta$:
\begin{eqnarray}{\cal L}_{\sigma\omega\rho\delta}
(\sigma,\omega _{_{\mu }},\vec{\rho }_{_{\mu }}, \vec{\delta}) ={
\cal L}_{\sigma\omega \rho}(\sigma,\omega _{_{\mu }},\vec{\rho
}_{_{\mu }})-U(\sigma)+ {\cal L}_{\delta}(\vec{\delta}),
\label{Lg}
\end{eqnarray}where $\cal L_{\sigma\omega \rho}$ is the linear part of
Lagrangian density without $\delta$-meson field\cite{Gl00},
$U(\sigma)=\frac{b}{3}m_{N}\left( g_{\sigma }\sigma \right)
^{3}+\frac{c}{4}\left( g_{\sigma }\sigma \right) ^{4}$ and $\cal
L_{\delta}$$(\vec{\delta})=g_{\delta } \bar {\psi}_{N} \vec{\tau
}_{N} \vec{\delta }\psi_ {N}+\frac{1}{2}\left(\partial _{\mu
}\vec{\delta}\partial ^{\mu}
\vec{\delta}-m_{\delta}\vec{\delta}^{2}\right)$ are the
$\sigma$-meson self-interaction term and the contribution of the
$\delta$-meson field, respectively. This Lagrangian density
(\ref{Lg}) contains the meson-nucleon coupling constants,
$g_{\sigma },~ g_{\omega },~g_{\rho },~g_{\delta}$, and also the
parameters of $\sigma$-field self-interacting terms, $b$ and $c$.
To examine the influence of the  meson field, $\delta$, on the
deconfinement phase transition characteristics, we use the model
parameter sets obtained in our recent work (see Ref.~\refcite{Al}
for details):
$a_{\sigma}=\left(g_{\sigma}/m_{\sigma}\right)^2=9.154$ fm$^2$,
$a_{\omega}=\left(g_{\omega}/m_{\omega}\right)^2=4.828$ fm$^2$,
$a_{\rho}=\left(g_{\rho}/m_{\rho}\right)^2=13.621$ fm$^2$,
$a_{\delta}=\left(g_{\delta}/m_{\delta}\right)^2=2.5$ fm$^2$,
$b=1.654\cdot10^{-2}$ fm$^{-1}$, $c=1.319\cdot10^{-2}$. If we
neglect the $\delta$ channel, we get $a_{\delta}=0$ and
$a_{\rho}=4.794$ fm$^{2}$. The standard QHD procedure allows to
obtain the energy density, $\varepsilon(n,\alpha)$, and pressure,
$P(n,\alpha)$, as a function of the baryon number density $n$ and
the asymmetry parameter $\alpha=(n_n-n_p)/n$.

    To describe the quark phase the MIT bag model is used, in which
the interactions between the $u,~d,~s$  quarks are taken in
one-gluon exchange approximation\cite{Far}. We choose $m_{u} = 5$
MeV, $m_{d} = 7$ MeV and $m_{s} = 150$ MeV for masses   , and
$B=100$ MeV/fm$^3$ for bag parameter and $\alpha_{s}=0.5$ for the
strong interaction constant.

    The chemical potentials of the constituents of the $npe$-plasma in
$\beta$-equilibrium are expressed through the two independent
potentials, $\mu_{b}^{(NM)}$ and $\mu_{el}^{(NM)}$, corresponding
to conserving of baryonic and electric charges:
\begin{equation}\label{muN}
\mu_{n}=\mu_{b}^{(NM)},~~~~\mu_{p}=\mu_{b}^{(NM)}-\mu_{el}^{(NM)},
~~~~\mu_{e}=\mu_{el}^{(NM)}.
\end{equation}
In this case, the pressure, $P_{NM}$, energy density,
$\varepsilon_{NM}$ and baryon number density, $n_{NM}$, are
functions of potentials, $\mu_{b}^{(NM)}$ and $\mu_{el}^{(NM)}$.
The particle species chemical potentials for $udse$-plasma in
$\beta$-equilibrium  are expressed through the chemical
potentials, $\mu_{b}^{(QM)}$ and $\mu_{el}^{(QM)}$:
\begin{equation}\label{muQ}
\mu_{u}=\frac{1}{3}\left(\mu_{b}^{(QM)}-2~\mu_{el}^{(QM)}\right),~
\mu_{d}=\mu_{s}=\frac{1}{3}\left(\mu_{b}^{(QM)}+\mu_{el}^{(QM)}\right),~
\mu_{e}=\mu_{el}^{(QM)}.
\end{equation}

In this case, the pressure, $P_{QM}$, energy density,
$\varepsilon_{QM}$ and baryon number density, $n_{QM}$, are
functions of chemical potentials $\mu_{b}^{(QM)}$ and
~$\mu_{el}^{(QM)}$.

The mechanical and chemical equilibrium conditions (Gibbs
conditions) for the mixed phase are:
\begin{equation}\label{Gb1}
\mu_{b}^{(QM)}=\mu_{b}^{(NM)}=\mu_{b},~~~~\mu_{el}^{(QM)}=\mu_{el}^{(NM)}=\mu_{el},
\end{equation}
\begin{equation}\label{Gb2}
P_{QM}(\mu_{b},~\mu_{el})=P_{NM}(\mu_{b},~\mu_{el}).
\end{equation}

We applied the global electrical neutrality condition for mixed
quark-nucleonic matter, according to Glendenning\cite{Gl00},
\begin{equation}\label{ch}
(1-\chi)\left(n_{p}-n_{e}\right)+\chi\left(\frac{2}{3}~n_{u}-\frac{1}{3}~n_{d}-\frac{1}{3}~n_{s}-n_{e}\right)=0.
\end{equation}
Here $\chi=V_{QM}/(V_{QM}+V_{NM})$ is the volume fraction of the
quark phase, where $V_{QM}$ and $V_{NM}$ are the volumes occupied
by the quark matter and nucleonic matter, respectively.

The baryon number density and energy density in the mixed phase
are:
\begin{equation}\label{n}
n=(1-\chi)\left(n_{p}+n_{n}\right)+\frac{1}{3}~\chi\left(n_{u}+n_{d}+n_{s}\right),
\end{equation}
\begin{equation}\label{eps}
\varepsilon=(1-\chi)\left(\varepsilon_{p}+\varepsilon_{n}\right)+\chi\left(\varepsilon_{u}+\varepsilon_{d}+\varepsilon_{s}\right)+\varepsilon_{e}.
\end{equation}

In case of $\chi=0$, the chemical potentials, $\mu_{b}^{N}$ and
$\mu_{el}^{N}$, corresponding to the lower threshold of the mixed
phase, are determined solving Eqs. (\ref{Gb2}) and (\ref{ch}).
This allows to find the lower boundary parameters, $P_{N}$,
$\varepsilon_{N}$ and $n_{N}$. Similarly, we calculate the upper
boundary values of mixed phase parameters, $P_{Q}$,
$\varepsilon_{Q}$ and $n_{Q}$, for $\chi=1$. The system of Eqs.
(\ref{Gb2}), (\ref{ch}), (\ref{n}) and (\ref{eps}) makes possible
to determine EOS of the mixed phase between this critical states.

Note, that in the case of an ordinary first-order phase transition
both nuclear and quark matter are assumed to be separately
electrically neutral, and at some pressure, $P_{0}$, corresponding
to the coexistence of the two phases, their baryon chemical
potentials are equal, i.e., $\mu _{NM}( {P_{0}}) = \mu _{QM}(
{P_{0}})$. Such a phase transition scenario is known as the phase
transition with constant pressure (MC).
\begin{table}[ph]
\tbl{Threshold parameters of the deconfinement phase transition
for both MC and GC scenarios with and without $\delta$ -meson
field } {\begin{tabular}{@{}ccccccc@{}} \toprule
Model & $n_{N}$ & $n_{Q}$& $P_{N}$ & $P_{Q}$ & $\varepsilon_{N}$ & $\varepsilon_{Q}$\\
& (fm$^{-3})$& (fm$^{-3})$& (MeV/fm$^{3})$& (MeV/fm$^{3})$&(MeV/fm$^{3})$& (MeV/fm$^{3})$\\
\colrule
GC with $\delta$      \hphantom{00}& \hphantom{0}0.241 & \hphantom{0}1.448 &\hphantom{0}16.9 &\hphantom{0}474.4&\hphantom{0}235.0 & 1889.3\\
GC without $\delta$   \hphantom{00}& \hphantom{0}0.260 & \hphantom{0}1.436 &\hphantom{0}18.0 &\hphantom{0}471.3&\hphantom{0}253.8 & 1870.8\\
MC with $\delta$      \hphantom{00}& \hphantom{0}0.475 & \hphantom{0}0.650 &\hphantom{0}93.3 &\hphantom{0}93.3 &\hphantom{0}503.3& 723.5\\
MC without $\delta$   \hphantom{00}& \hphantom{0}0.551 & \hphantom{0}0.717 &\hphantom{0}121.3 &\hphantom{0}121.3&\hphantom{0}593.4 & 810.0\\
\botrule
\end{tabular} \label{ta1}}
\end{table}

\begin{figure}[t]
\begin{center}
  \begin{minipage} {0.47\linewidth}
   \begin{center}
\centerline{\psfig{file=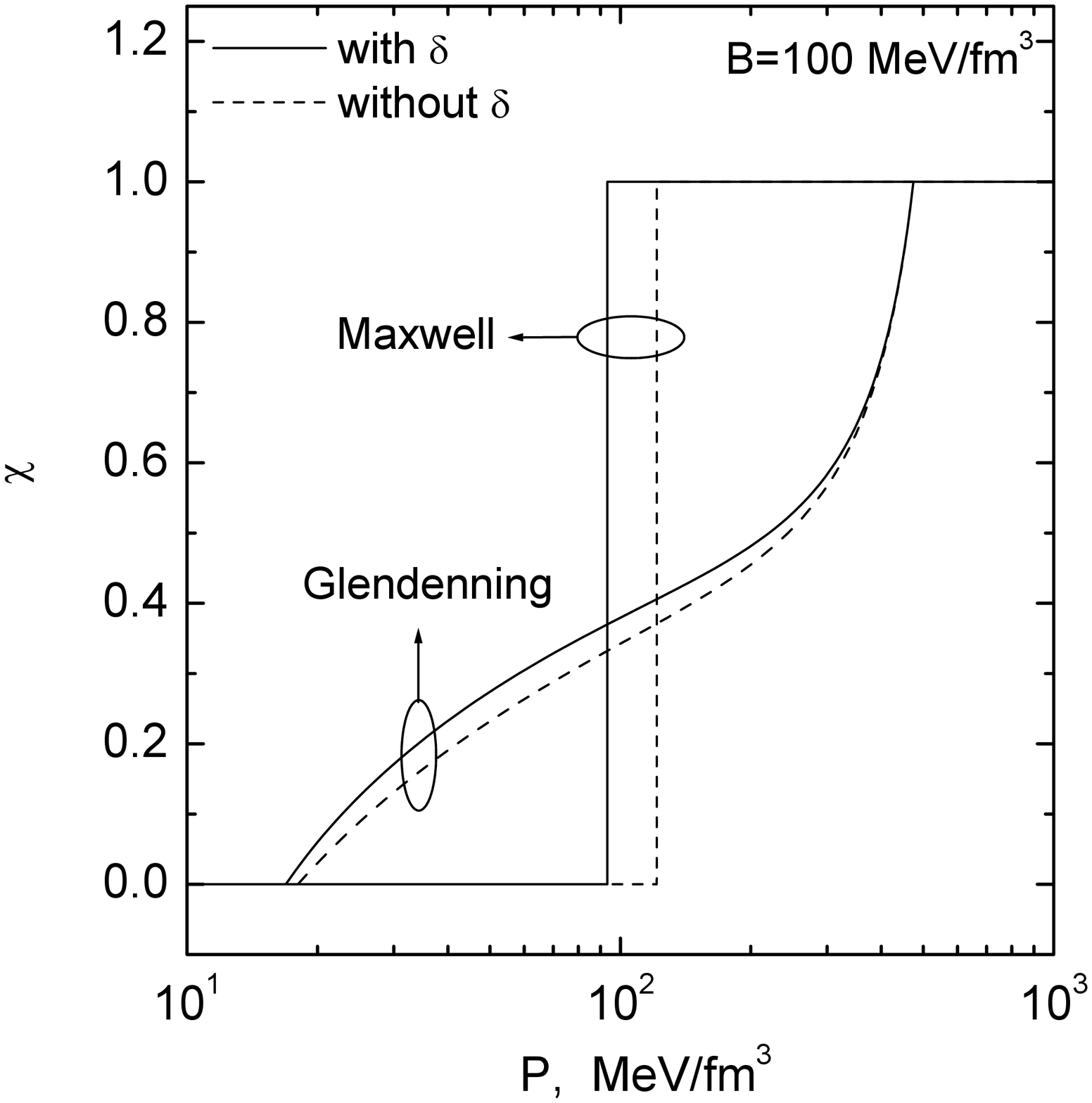,width=5.4 cm}} \vspace*{8pt}
 \caption {The volume fraction, $\chi$, as
a function of pressure, $P$, for MC and GC scenarios with and
without $\delta$-meson field.}
    \end{center}
  \end{minipage}\label{Fg1} \hfil\hfil
 \begin{minipage} {0.47\linewidth}
   \begin{center}
\centerline{\psfig{file=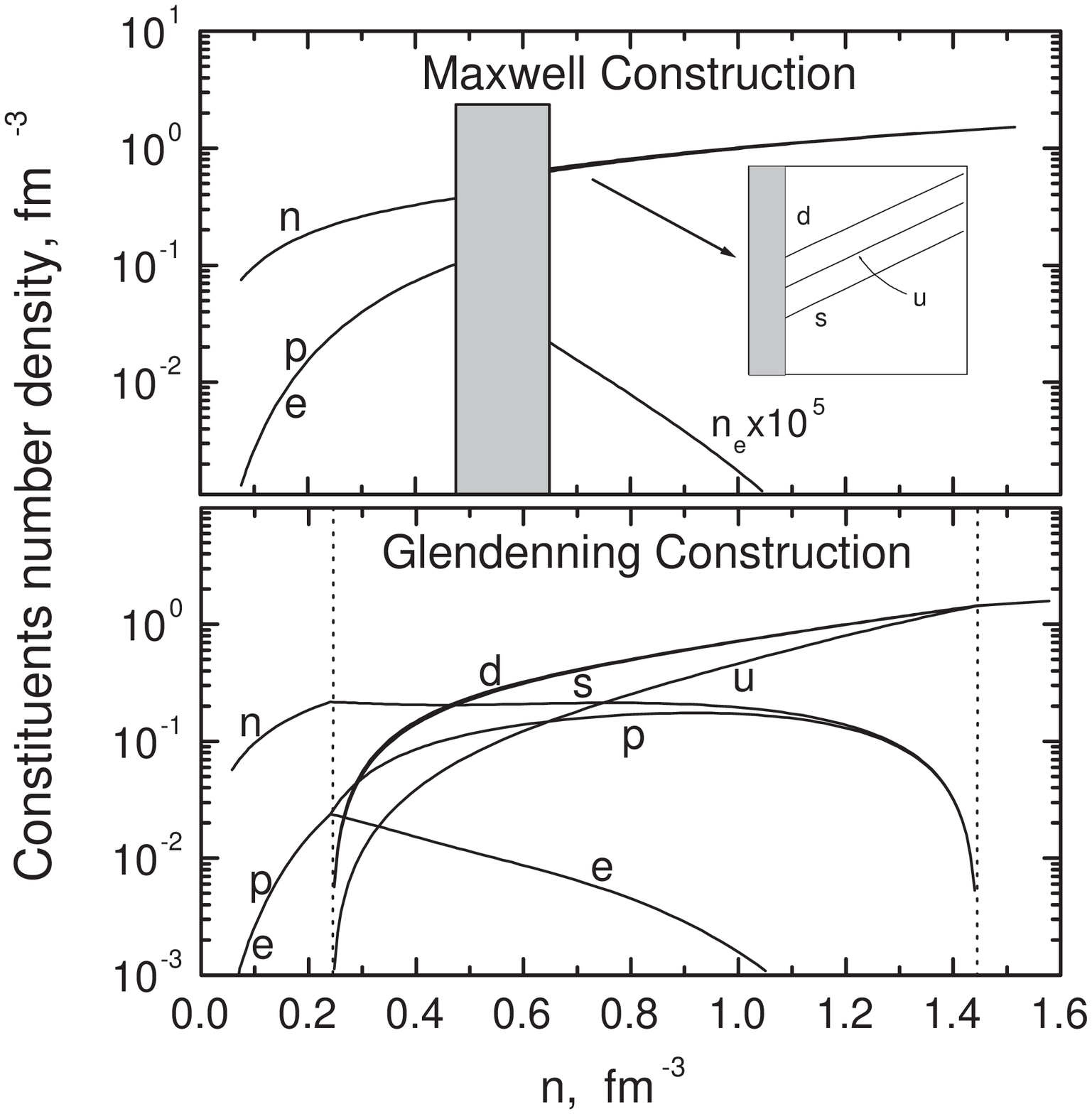,width=5.4 cm}} \vspace*{8pt}
   \caption {Constituents composition as a function of baryon number density, $n$, for MC and GC scenarios with $\delta$-meson field.}
    \end{center}
  \end{minipage}\label{Fg2}
 \end{center}
\end{figure}

Table 1 represents the threshold parameter sets of the quark phase
transition with and without $\delta$-meson field. It is shown that
the presence of $\delta$-field alters the threshold
characteristics of phase transition.

In Fig.1 we plot the quark matter volume fraction, $\chi$, as a
function of pressure, $P$. The solid lines correspond to both
scenarios MC and GC with $\delta$-meson field, while the dashed
lines correspond to this scenarios without $\delta$-meson field.

In Fig.2 we plot the particle species number densities as a
function of the baryon density, $n$. the upper panel corresponds
to the MC scenario and the lower one corresponds to the GC
scenario. The MC scenario leads to appearance of a discontinuity.
The charge neutral nucleonic matter at the baryon density
$n_{N}=0.475$ fm$^{-3}$ coexists with the charge neutral quark
matter at the baryon density $n_{Q}=0.650$ fm$^{-3}$. Thus, the
density range $n_{N}<n<n_{Q}$ is forbidden. In GC scenario the
quarks appear at the critical density $n_{N}=0.241$ fm$^{-3}$ and
the hadronic matter completely disappears at $n_{Q}=1.448$
fm$^{-3}$, where the pure quark phase occurs.

\section{Conclusion}
We show that the scalar – isovector $\delta$-meson field inclusion
leads to the increase of the EOS stiffness of nuclear matter due
to the splitting of proton and neutron effective masses, and also
due to the increase of the asymmetry energy. The presence of
$\delta$-meson field alters the threshold characteristics of
deconfinement phase transition. The lower threshold parameters,
$n_N$, $\varepsilon_N$, $P_N$, for GC scenario  decrease,
meanwhile the upper ones, $n_Q$, $\varepsilon_Q$, $P_Q$, slowly
increase. In case of MC scenario the coexistence pressure, $P_0$,
decreases. These alterations of the phase transition parameters
can lead to the corresponding alterations of structural and
integral characteristics of neutron stars with quark degrees of
freedom.

\section*{Acknowledgments}
We wish to acknowledge Prof. Yu.L. Vartanyan and Dr. G.S. Hajyan
for fruitful discussions. This research is supported by the
Ministry of Education and Sciences of the Republic of Armenia
under grant 2008-130.
%\begin{thebibliography}{000} %for 3 digits
%\begin{thebibliography}{00}  %for 2 digits

\end{document}